\documentclass[useAMS,usenatbib]{mn2e}

\usepackage{graphicx}
\usepackage{natbib}
\usepackage{amsmath}
\usepackage{deluxetable}

\def\lesssim{\mathrel{\hbox{\rlap{\hbox{\lower5pt\hbox{$\sim$}}}\hbox{$<$}}}}
\def\gtrsim{\mathrel{\hbox{\rlap{\hbox{\lower5pt\hbox{$\sim$}}}\hbox{$>$}}}}

\def\apjl{ApJL}

\def\apjs{ApJS}
\def\aap{A\&A}

\usepackage{color}

\title[Rapid Eccentricity Oscillations in Hierarchical Triples]{Rapid
Eccentricity Oscillations and the Mergers of Compact Objects in
Hierarchical Triples}

\author[Antognini et al.]{Joe M.~Antognini$^1$, Benjamin
J.~Shappee$^1$, Todd A.~Thompson$^{1, 2}$, Pau Amaro-Seoane$^3$
\\
$^1$ Department of Astronomy, The Ohio State University, Columbus, Ohio
43210, USA\\
$^2$ Center of Cosmology and Astro-Particle Physics, The Ohio State
University, Columbus, Ohio 43210, USA\\
$^3$ Max Planck Institut f\"ur Gravitationsphysik
(Albert-Einstein Institut), D-14476 Potsdam, Germany\\
E-mail: antognini@astronomy.ohio-state.edu}

\begin{document}

\maketitle

\begin{abstract}

Kozai-Lidov (KL) oscillations can accelerate compact object mergers via
gravitational wave (GW) radiation by driving the inner binaries of
hierarchical triples to high eccentricities.  We perform direct three-body
integrations of high mass ratio compact object triple systems using
\textsc{Fewbody} including post-Newtonian terms.  We find that the inner
binary undergoes rapid eccentricity oscillations (REOs) on the timescale of
the outer orbital period which drive it to higher eccentricities than secular
theory would otherwise predict, resulting in substantially reduced merger
times.  For a uniform distribution of tertiary eccentricity ($e_2$),
$\sim40$\% of systems merge within $\sim1-2$ eccentric KL timescales whereas
secular theory predicts that only $\sim$20\% of such systems merge that
rapidly.  This discrepancy becomes especially pronounced at \emph{low} $e_2$,
with secular theory overpredicting the merger time by many orders of
magnitude.  We show that a non-negligible fraction of systems have
eccentricity $>0.8$ when they merge, in contrast to predictions from secular
theory.  Our results are applicable to high mass ratio triple systems
containing black holes or neutron stars.  In objects in which tidal effects
are important, such as white dwarfs, stars, and planets, REOs can reduce the
tidal circularization timescale by an order of magnitude and bring the
components of the inner binary into closer orbits than would be possible in
the secular approximation.

\end{abstract}

\section{Introduction}
\label{sec:intro}

Hierarchical triple systems are common \citep{raghavan10} and exhibit
dynamics that are qualitatively different from binary systems
\citep{poincare92}.  For example, if the tertiary is highly inclined with
respect to the inner binary, it induces slow oscillations of the orbital
parameters of the inner binary.  In particular, the eccentricity of the
inner binary oscillates between a minimum and maximum value ($e_{\rm{max}}
= \sqrt{1 - 5/3 \cos^2 i}$ in the limit of a test particle secondary when
the three-body Hamiltonian is expanded to quadrupole order) over the
timescale \citep[e.g.,][]{holman97, blaes02}
\begin{multline}
\label{eq:kozai_timescale}
t_{\textrm{KL}} \sim 1.3 \times 10^5 \, \textrm{yr} \left( \frac{m_1 +
m_2}{2 \times 10^6 \, M_{\odot}} \right)^{-1/2} \left( \frac{a_1}{10^{-2}
\, \textrm{pc}} \right)^{3/2} \\
\times \left( \frac{m_1 + m_2}{2 m_2} \right) \left( \frac{a_2 / a_1}{10}
\right)^3 \left( 1 - e_2^2 \right)^{3/2}.
\end{multline}
These oscillations are known as Kozai-Lidov oscillations \citep{kozai62,
lidov62}.

Kozai-Lidov oscillations have found application in a wide variety of
astrophysical systems including the orbits of asteroids in the Solar System
\citep{kozai62}, the orbits of artificial satellites around planets in the
Solar System \citep{lidov62}, as a formation channel for hot Jupiters
\citep[e.g.,][]{wu03, fabrycky07, wu07}, and as a formation channel for
blue stragglers \citep{perets09}.

Kozai-Lidov cycles can drive the inner binary in some hierarchical triple
systems to merger via gravitational wave emission \citep{blaes02,
miller02a}.  If the inner binary consists of two white dwarfs,
\citet{thompson11} showed that these mergers occur rapidly enough to
potentially explain the Type Ia supernova rate.  \citet{katz12}
demonstrated that in a non-negligible fraction of systems, perturbations to
the secular Kozai-Lidov oscillations can drive the inner binary to collide
head-on, rather than coalescing due to gravitational radiation after tidal
capture \citep[see also][]{prodan13}.  \citet{hamers13} provide a more
detailed discussion of the rates of such collisions by accounting for the
evolution of the inner binary on the main sequence.

In addition to WD-WD mergers, Kozai-Lidov oscillations have also been
studied as a mechanism to drive other compact objects to rapid merger.
Neutron star-neutron star and neutron star-black hole mergers have been
proposed as engines of short gamma-ray bursts \citep{paczynski86,
ruffert95, ruffert96, ruffert99, janka99}, and such mergers may also be
expedited by Kozai-Lidov oscillations \citep{thompson11}.  Several authors
have studied mergers of stellar-mass black holes, particularly in globular
clusters, to determine if they can efficiently grow to intermediate-mass
black holes \citep[IMBHs; e.g.,][]{miller02a, wen03, gultekin04,
aarseth12}.  Hierarchical triples of supermassive black holes (SMBHs) may
also merge quickly as a result of Kozai-Lidov oscillations \citep{blaes02,
hoffman07, amaro-seoane10}.  These effects can also produce interesting
gravitational wave signatures from stellar mass binaries in orbit around
one or more SMBHs \citep[e.g.,][]{antonini12, bode13}.

Given its general nature, the physics behind the Kozai-Lidov mechanism has
come under broader study in the past several years.  Until recently, almost
all work exploring it has employed the secular approximation, which assumes
that any changes to the orbital parameters of the system are slow compared
to the orbital period of the outer binary.  The Hamiltonian is expanded in
powers of the ratio of the semi-major axis of the inner binary to the
semi-major axis of the outer binary $(a_1/a_2)$, typically to quadrupole
order, $(a_1/a_2)^2$.  \citet{krymolowski99} and \citet{ford00, ford00err}
derived the equations of motion to octupole order, $(a_1/a_2)^3$
\citep[see][]{naoz13a}.  \citet{lithwick11} and \citet{katz11} explored the
implications of these equations and showed that the octupole-order terms
can lead to substantially larger eccentricities of the inner binary.  This
so-called eccentric Kozai mechanism (EKM) has dramatically expanded the
parameter space in which mergers and other interesting dynamics can occur
\citep[e.g.][]{naoz12, shappee13}.

It is becoming increasingly evident, however, that the secular
approximation can fail in certain circumstances.  \citet{antonini12} found
that in extreme-mass-ratio systems, eccentricities change rapidly compared
to the period of the tertiary if the tertiary is in an eccentric orbit
\citep[this behavior can also be seen in][]{antonini10}.  \citet{bode13}
found that in a more general set of systems, the eccentricity of the inner
binary varies on the timescale of the orbit of the tertiary.  Recently,
\citet{katz12} found that these rapid variations can lead to collisions of
WD-WD binaries if the tertiary is at very high
inclination.\footnote{\citet{katz12} distinguish between ``head-on
collisions,'' in which two objects merge without substantial tidal
interaction, and ``collisions,'' in which two objects merge with or without
previous tidal interaction.  We use ``collision'' to refer exclusively to
mergers without tidal interaction.  Any event in which the two objects
undergo substantial tidal interaction before combining is termed a
``merger'' in this paper.}  Finally, \citet{seto13} examined the impact of
these rapid fluctuations on gravitational wave astronomy.  

In this paper we revisit earlier calculations of the merger times of
compact objects by \citet{blaes02} and \citet{hoffman07}.  We extend
these works by directly integrating the equations of motion of the
three-body system and including post-Newtonian (PN) force terms up to order
3.5 to account for GR effects.  We show that motion of the tertiary on its
orbit (even in relatively low eccentricity orbits) leads to rapid
eccentricity oscillations (REOs) in the inner binary and we quantify the
importance of these oscillations.  Our goal is to better understand
the effect of the eccentric Kozai-Lidov mechanism and
non-secular effects on the merger time distribution and dynamics of compact
object binaries.  In systems with
tertiaries in low eccentricity orbits we find that the
double-orbit-averaged secular approximation fails by predicting merger
times many orders of magnitude longer than those of the direct three-body
integration.

This paper is structured as follows.  In \S \ref{sec:methods} we describe
our numerical methods and characterize the accuracy of our integration (see
also the Appendix).  In \S \ref{sec:reo} we describe the breakdown of the
secular approximation in calculating the eccentricity of the inner binary.
In \S \ref{sec:mergetime} we demonstrate one regime in which this breakdown
of the secular approximation leads to catastrophic failure, namely in
predicting the merger times of compact objects.  We conclude and discuss
a number of applications in \S \ref{sec:conclusions}.

As this manuscript was being completed, \citet{antonini13} presented
similar results on the breakdown of the secular approximation, the delay
time distribution, and the eccentricity distribution of compact object
binaries at merger.

\section{Numerical methods \& Setup}
\label{sec:methods}

We numerically evolve triple systems with the open source \textsc{Fewbody}
suite \citep{fregeau04}.  \textsc{Fewbody} is designed to compute the
dynamics of hierarchical systems of small numbers of objects ($N \lesssim
10)$ either in scattering experiments or in bound systems.  The underlying
integrator for the \textsc{Fewbody} suite is the GNU Scientific Library
ordinary differential equations library \citep{gough09}.  By default
\textsc{Fewbody} uses eighth-order Runge-Kutta Prince-Dormand integration
with adaptive time steps.  It is straightforward to modify \textsc{Fewbody}
to use any of the other roughly half-dozen integration algorithms supported
by GSL.\footnote{GSL also supports an additional five integration
algorithms, but these require the calculation of the Jacobian.  When
post-Newtonian terms are included this becomes nontrivial to implement.} In
our experience the choice of integration algorithm does not affect the
results since the adaptive steps force the size of the error to be within
the same target value regardless of the algorithm used.  All results in
this paper were obtained using the default eighth-order Runge-Kutta
Prince-Dormand algorithm.  To incorporate relativistic effects, we have
included post-Newtonian (PN) terms up to order 3.5 in the integration.
Details of energy conservation, gravitational radiation, and a comparison
to secular calculations are provided in Appendix \ref{sec:appendix}.  These
additions to \textsc{Fewbody} and a direct application to the formation of
gravitational-wave sources for ground-based detectors will be presented in
more detail in \citet{amaro-seoane13c}.

Throughout this paper we use $m_1$ and $m_2$ to refer to the masses of the
objects in the inner binary, and $m_3$ to refer to the mass of the
tertiary.  For other quantities the subscript `1' refers to the inner
binary and the subscript `2' refers to the outer binary.  The semi-major
axis is represented by $a$, the eccentricity by $e$, the argument of
periapsis by $g$, and the mutual inclination by $i$.

\section{Rapid eccentricity oscillations}
\label{sec:reo}

Most studies of three-body dynamics have employed the secular approximation
in which any changes to the orbital parameters of either orbit are assumed
to be slow compared to the orbital period of both orbits.  Such models
cannot account for any changes that occur on more rapid timescales, and it
is implicitly assumed that if such changes do occur, their effect would be
negligible.

We find that over a broad region of parameter space, the inner binaries in
triple systems undergo oscillations in eccentricity (or, equivalently,
angular momentum) on the timescale of the outer orbital period (``rapid
eccentricity oscillations,'' REOs).  REOs are typically small
and do not affect the dynamics of the triple system for almost all of its
evolution.  But when the inner binary is already at high eccentricity, as
during an eccentric Kozai cycle, the magnitude of the oscillations in
angular momentum becomes comparable to the total angular momentum of the
inner binary.  REOs can then drive the inner binary to rapid merger.

The existence of REOs was predicted by \citet{ivanov05}, who found that the
amplitude of the change in angular momentum during an oscillation is
\begin{equation}
\label{eq:reo}
\frac{\Delta L}{\mu} = \frac{15}{4} \frac{m_3}{m_1 + m_2} \cos i_{\min}
\left( \frac{a_1}{a_2} \right)^2 \sqrt{G m_3 a_2},
\end{equation}
where $\mu$ is the reduced mass of the inner binary and $i_{\min}$ is the
minimum mutual inclination between the two orbits during a Kozai-Lidov
cycle.\footnote{See Appendix B of \citealt{ivanov05} for the complete
derivation.  Note that in \citealt{ivanov05}, $\Delta L$ refers to the
change in the specific angular momentum from the average value to the
maximum value.  This quantity therefore differs from our $\Delta L$ by a
factor of $\mu/2$.}  Equation~\ref{eq:reo} can also be written as a change
in eccentricity, although this form of the equation is somewhat more
cumbersome:
\begin{multline}
\label{eq:reo_ecc}
\Delta e_1 = -e_1 + \\
\sqrt{1 - \left[ \sqrt{1 - e_1^2} + \frac{15}{4} \left( \frac{m_3}{m_1 +
m_2} \right)^{\frac{3}{2}} \cos i_{\min} \left( \frac{a_1}{a_2}
\right)^{\frac{3}{2}} \right]^2}.
\end{multline}
These equations are only accurate near the eccentricity maximum of
a Kozai-Lidov cycle.

\begin{figure*}
\centering
\includegraphics[width=16cm]{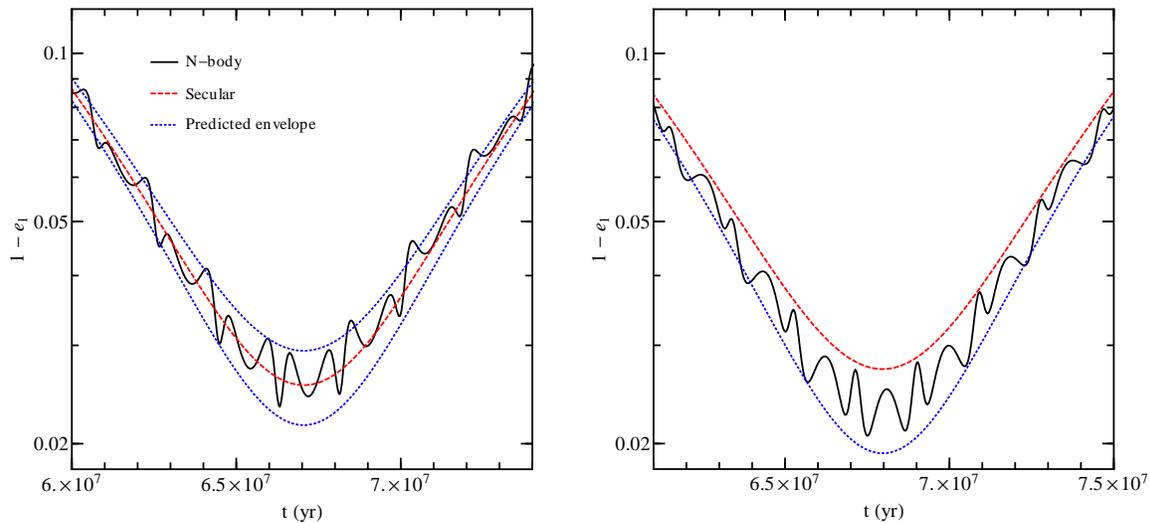}

\caption{Systems exhibiting REOs.  The eccentricity of the inner binary
during a Kozai-Lidov cycle as calculated by direct three-body integration
(black solid line) and as calculated in the secular approximation (red
dashed line) for $g_1 = g_2 = 0^{\circ}$ (left panel) and $g_1 - g_2 =
90^{\circ}$ (right panel).  PN terms are not included.  The secular and
three-body calculations match on average in the left panel, but the
three-body calculation exhibits oscillations in $e_1$.  In the right panel,
the secular calculation correctly predicts the minimum eccentricity, but
the REOs in the three-body calculation push the binary exclusively to
higher eccentricities.  Blue dotted lines show the amplitude of the REOs
predicted by Equation \ref{eq:reo}.  The period of the REOs is twice the
period of the outer binary.  The asymmetry in the period of the
oscillations is due to fact that the tertiary is on an eccentric orbit
($e_2 = 0.2$).  The initial conditions of the system are presented in Table
\ref{tbl:params} but with $g_1$ and $g_2$ fixed as stated above.}
\label{fig:reo}

\end{figure*}

Our numerical experiments are in agreement with \citet{ivanov05}.  We show
in Figure~\ref{fig:reo} the evolution of two example systems exhibiting
REOs (see Table \ref{tbl:params}).  The two systems are identical except
that the system in the left panel begins with $g_1 = g_2 = 0^{\circ}$ and
the right panel begins with $g_1 - g_2 = 90^{\circ}$.  To demonstrate that
REOs are a non-relativistic phenomenon, we have suppressed PN terms in this
figure.

\begin{table*}
\centering

\caption{Initial conditions for triple systems studied in this paper.
Throughout this paper $g$ refers to the argument of periapsis and $i$
refers to the mutual inclination.}\label{tbl:params}

\begin{tabular}{cccccccccc}

\hline

$m_1$ &
$m_2$ &
$m_3$ &
$a_1$ &
$a_2$ &
$e_1$ &
$e_2$ &
$g_1$ &
$g_2$ &
$i$
\\

\hline

$10^7$ M$_{\odot}$ &
$10^5$ M$_{\odot}$ &
$10^7$ M$_{\odot}$ &
1 pc &
20 pc &
0.1 &
0.1--0.8 &
0--360$^{\circ}$ &
0--360$^{\circ}$ &
80$^{\circ}$
\\

\hline

\end{tabular}
\end{table*}

Intuitively, REOs can be understood as similar to a Kozai-Lidov
oscillation in miniature.  In the double-orbit-averaged approximation, the
Kozai-Lidov mechanism occurs due to the fact that the outer orbit exerts a
stronger force on the inner orbit at the line of nodes than at other
regions of the outer orbit.  But because in reality the outer orbit is a
point mass in motion rather than a continuous hoop of matter, this force is
strongest along the line of nodes as the tertiary is actually passing
through the line of nodes.  For weak Kozai-Lidov oscillations, the driving
force contributed during any single orbit is small, so there is only a
gradual change in the eccentricity of the inner orbit and any rapid
eccentricity oscillations are negligible.  However, during a strong
eccentricity maximum, the inner orbit has lost nearly all of its angular
momentum and is therefore extremely sensitive to torquing.

This implies that the arguments of periapsis of the inner and outer orbits
determine the direction of the oscillation.  If the apsides are aligned
with the line of nodes (as in the left panel of Figure~\ref{fig:reo}), the
eccentricity will be driven to higher values relative to the secular
calculation when the tertiary passes through periapsis and to lower values
when the tertiary passes through apoapsis.  If the apsides are $90^{\circ}$
from the line of nodes, however, the eccentricity will be exclusively
driven to higher values relative to the secular calculation while the
amplitude of the oscillations will remain fixed (as in the right panel of
Figure~\ref{fig:reo}).

Although oscillations in the orbital elements on the timescale of the
period of the outer orbit were first predicted by \citet{soderhjelm75}, an
explicit formula for the change in angular momentum was first derived by
\citet{ivanov05}.  Moreover, these oscillations were not confirmed by
three-body integrations until \citet{bode13}, who found them in the
test-particle limit, and \citet{antonini12}, who found them in the
equal-mass case.  \citet{katz12} further explored these oscillations in the
context of WD-WD collisions.  They argued that these oscillations are
fundamentally a stochastic phenomenon, but only examined systems in which
the inclination of the tertiary was near the Kozai ``pole'' of $i \sim
93^{\circ}$, where certain terms in the Hamiltonian formally diverge at
quadrupole order and Kozai-Lidov oscillations become extremely strong
\citep{miller02a}.  Although a complete analytic treatment of REOs is
beyond the scope of this paper, our results suggest that at lower
inclinations they could be modelled analytically.  As we discuss in Section
\ref{subsec:population}, we only examine REOs in inner binaries on prograde
orbits.

\section{Effect on merger times}
\label{sec:mergetime}

REOs are important when $1-e_1 \sim 0$ and nearly all of the angular
momentum in the inner orbit has been transferred to the outer orbit. Here,
fluctuations in the angular momentum given by Equation~\ref{eq:reo} become
comparable to the total angular momentum in the inner orbit itself.  REOs
then can have a substantial impact on the long-term dynamics.  This is
especially true if relativistic effects are important because the PN terms
are strong functions of distance and thus a small change in the distance at
periapsis dramatically changes their strength.  A secular calculation does
not account for these effects and will overpredict the merger time, in some
cases by many orders of magnitude. There are therefore certain regions of
parameter space in which the double-orbit-averaging approximation fails.

\subsection{Test case}
\label{subsec:testcase}

\begin{figure}
\centering
\includegraphics[width=8cm]{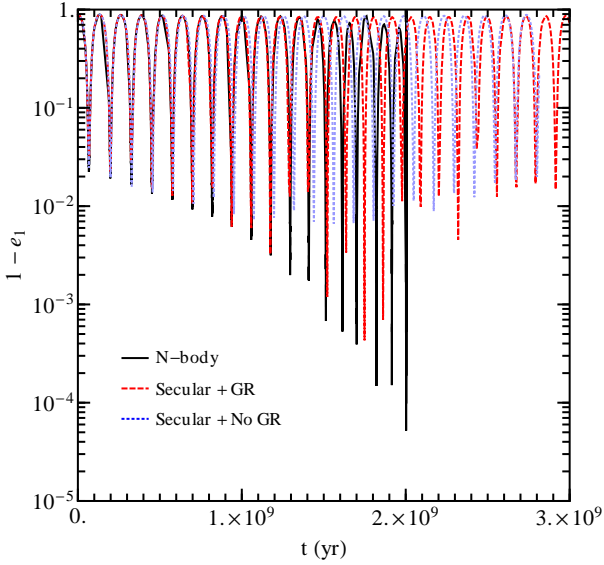}

\caption{The impact of REOs on the evolution of the inner binary of a
hierarchical triple.  We show the direct three-body integration (solid
black line) and the calculation in the secular approximation (red dashed
line).  To illustrate the importance of relativistic terms at high
eccentricities we also show the direct three-body integration without any
PN terms (blue dotted line).  Although the direct integration matches
closely with the two secular calculations during the first Kozai-Lidov
cycle, the calculations diverge thereafter.  The secular calculation
predicts that the system only reaches a maximum eccentricity of $1-e_1 \sim
10^{-3}$ in the time period shown, whereas the direct integration predicts
that the inner binary is driven to sufficiently high eccentricities to
merge after $\sim 2 \times 10^9$ yr.} \label{fig:example}

\end{figure}

The importance of REOs for the long-term evolution of an example system is
illustrated in Figure~\ref{fig:example}.  The secular calculation (red
dashed line) closely matches the three-body integration performed by
\textsc{Fewbody} (black line) during the first Kozai-Lidov cycle, but
afterwards they begin to diverge.  In this particular case, the
eccentricity of the inner binary in the three-body integration increases to
$1-e_1 < 10^{-4}$, whereas in the secular calculation it only reaches
$1-e_1 \sim 10^{-3}$.  As a consequence, the three-body integration
predicts the inner binary to merge within one eccentric-Kozai-Lidov
timescale whereas the secular calculation predicts that the inner binary
will effectively never merge.  To demonstrate the importance of resonant
post-Newtonian eccentricity excitation \citep{naoz13b} we additionally show
the same three-body calculation without any PN terms (blue dotted line).
We find that without relativistic effects the inner binary gets excited to
much lower eccentricities, in agreement with \citet{naoz13b}.

\begin{figure*}
\centering
\includegraphics[width=16cm]{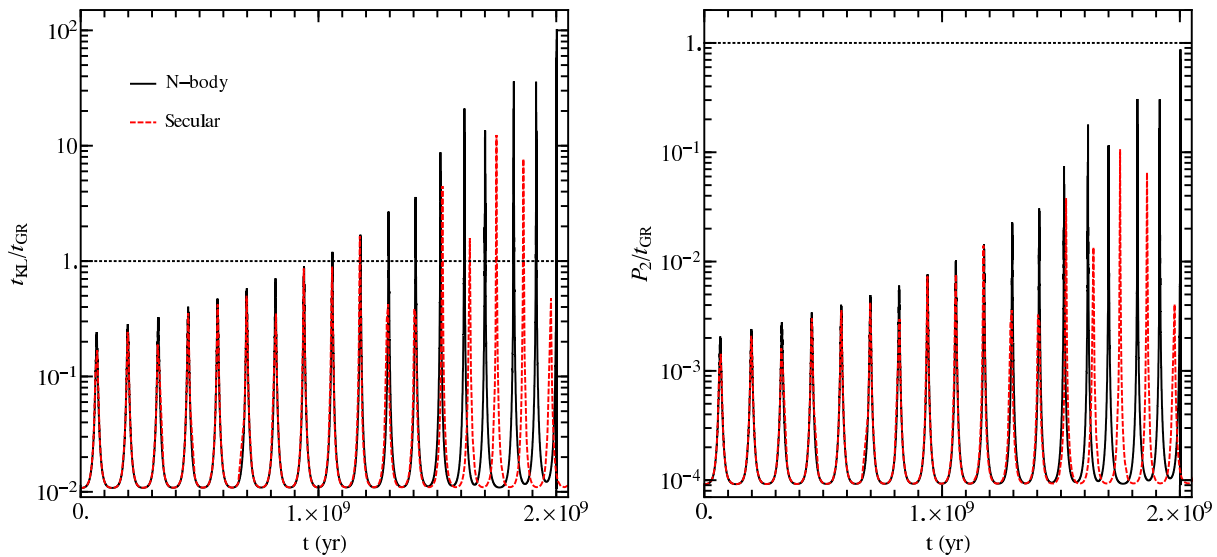}

\caption{Timescales of the system presented in Figure~\ref{fig:example}
calculated with direct integration (black line) and in the secular
approximation (dotted red line).  \emph{Left panel.}  The ratio between the
Kozai-Lidov timescale (Eq.~\ref{eq:kozai_timescale}) and the GR precession
timescale (Eq.~\ref{eq:gr}) of the inner binary.  During the final several
Kozai-Lidov cycles the GR precession timescale becomes shorter than the
Kozai-Lidov timescale.  Although the Kozai-Lidov mechanism is detuned, REOs
are sufficient to restore the system to lower eccentricities and continue
the Kozai-Lidov cycle.  \emph{Right panel.}  The ratio between the outer
period and the GR precession timescale.  Once the GR precession timescale
of the inner binary becomes comparable to the outer orbital period, the two
orbits completely decouple and gravitational radiation drives the inner
binary to merge.} \label{fig:timescales}

\end{figure*}

During the eccentric phase of the Kozai-Lidov cycles the GR precession
timescale, $t_{\rm{GR}}$, shortens since $e_1$ approaches unity
\citep[e.g.,][]{blaes02}:
\begin{equation}
\label{eq:gr}
t_{\textrm{GR}} \sim 2.3 \times 10^6 \, \textrm{yr} \left( \frac{m_1 +
m_2}{2 \times 10^6 \, M_{\odot}} \right)^{-3/2} \left( \frac{a_1}{10^{-2}
\, \textrm{pc}} \right)^{5/2} \left( 1 - e_1^2 \right).
\end{equation}
If the eccentricity becomes sufficiently large, as in the final Kozai-Lidov
cycles of the system presented in Figure~\ref{fig:example}, $t_{\rm{GR}}$
can become shorter than the Kozai-Lidov timescale, $t_{\rm{KL}}$
(Eq.~\ref{eq:kozai_timescale}).  This will ordinarily not suppress the
Kozai-Lidov mechanism because at high eccentricity the inner binary
requires only very small torques to change its eccentricity.
\citet{bode13} therefore introduce the \emph{instantaneous} Kozai-Lidov
timescale, $t_{\rm{KL, inst}}$ as the timescale for the inner binary to
change its angular momentum by order unity.  $t_{\rm{KL, inst}}$ is related
to $t_{\rm{KL}}$ by
\begin{equation}
\label{eq:tklinst}
t_{\rm{KL, inst}} \sim \sqrt{1 - e_1^2} \, t_{\rm{KL}}
\end{equation}
up to constant factors of order unity.  If $t_{\rm{KL, inst}}$ exceeds
$t_{\rm{GR}}$ the Kozai-Lidov cycles are ``detuned,'' and the Kozai-Lidov
mechanism will be suppressed \citep{holman97}.  This kind of detuning
occurs in the regime in which the secular approximation is valid.  The
detuning which occurs in the system presented in \ref{fig:example} does not
occur in this regime, however.  In this system $t_{\rm{GR}}$ becomes
shorter than $P_2$ before it becomes shorter than $t_{\rm{KL, inst}}$.
When this occurs, it is impossible for the outer binary to exert any
secular influence on the inner binary.  At this point, the inner binary
decouples from the outer binary and gravitational radiation drives the
inner binary to rapid merger because the inner binary is at high
eccentricity.  Hence the Kozai-Lidov mechanism is detuned as a consequence
of the breakdown of the secular approximation.  To emphasize that
$t_{\rm{KL}}$ is not relevant for determining if Kozai-Lidov cycles will be
detuned in the middle of a Kozai-Lidov cycle, we show in the left panel of
Figure~\ref{fig:timescales} the ratio between $t_{\rm{KL}}$ and
$t_{\rm{GR}}$ for the system presented in Figure~\ref{fig:example}.  During
the final Kozai-Lidov cycles $t_{\rm{GR}}$ becomes much shorter than
$t_{\rm{KL}}$, but is driven back to longer timescales by REOs before the
inner binary can merge by gravitational radiation.  The right panel shows
the ratio between $P_2$ and $t_{\rm{GR}}$ for the same system.  Once this
ratio reaches unity, the inner binary decouples from the outer binary and
merges via gravitational radiation.

\begin{figure}
\centering
\includegraphics[width=8cm]{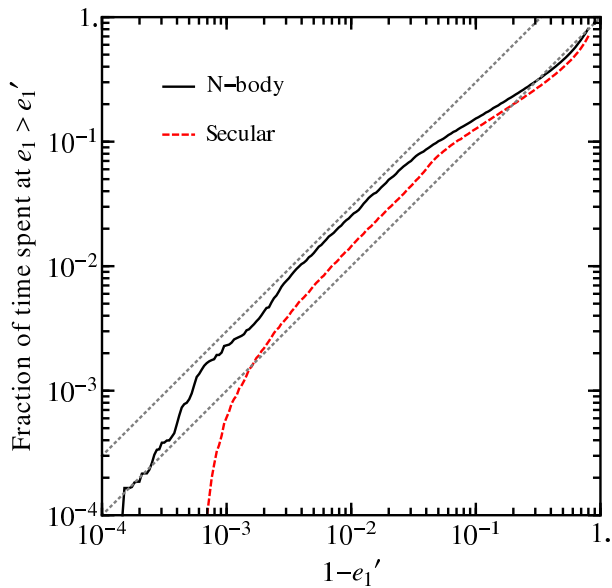}

\caption{The fraction of time that the system presented in
Figure~\ref{fig:example} spends at eccentricities greater than any given
eccentricity in the direct integration (solid line) and in the secular
approximation (red dashed line).  For comparison we also present the line
$y=x$ and $y=3x$ (dotted lines).  At low eccentricities, the fraction of
time that a hierarchical triple undergoing Kozai-Lidov cycles spends at
eccentricities greater than $e_1$ is approximately $1-e_1$.  REOs drive the
system to spend more time at higher eccentricities.  If eccentric
Kozai-Lidov oscillations are also present, as in this case, the fraction of
time spent at higher eccentricities is slightly larger, but is always
within a factor of a few of $1-e_1$.  Although REOs are only important at
high eccentricities, their effect during these brief periods drastically
changes the overall evolution of the system.} \label{fig:etimefrac}

\end{figure}

We emphasize that REOs are only important when the eccentricity is large.
REOs therefore only affect the dynamics during a small fraction of the
system's lifetime.  We illustrate in Figure~\ref{fig:etimefrac} the
fraction of time that the system spends at high eccentricity.  The
fluctuations in the angular momentum of the inner binary become 10\% of the
total angular momentum of the inner binary when the inner binary reaches an
eccentricity of $\sim$0.9.  From Figure~\ref{fig:etimefrac}, REOs are
therefore non-negligible for only $\sim$10\% of the system's lifetime.  The
secular approximation is therefore valid $\sim$90\% of the time.
Nevertheless, as illustrated in Figure~\ref{fig:example} and in the next
subsection, these short periods in which the secular approximation fails
dramatically influence the evolution of the system and lead to a sharp
divergence from the secular predictions because of the strong eccentricity
dependence of the GR terms.

\subsection{Population study}
\label{subsec:population}

Here we compare the merger times of a variety systems calculated in both
the secular approximation and in the full three-body integration.  We fix
the masses of the SMBHs ($m_1 = 10^7$ M$_{\odot}$, $m_2 = 10^5$
M$_{\odot}$, $m_3 = 10^7$ M$_{\odot}$), the semi-major axes ($a_1 = 1$ pc,
$a_2 = 20$ pc), the initial eccentricity of the inner binary ($e_1 = 0.1$),
the inclination of the tertiary ($i = 80^{\circ}$), and the arguments of
periapsis ($g_1 = 0^{\circ}$, $g_2 = 90^{\circ}$).  The initial mean
anomalies are chosen randomly.  The eccentricity of the tertiary, $e_2$, is
systematically varied from $e_2 = 0.1$ to 0.85 in steps of 0.001.  (Systems
at $e_2 \gtrsim 0.85$ are unstable and few systems with $e_2 \lesssim 0.1$
ever merge.)  Note that we do not choose our masses to model any specific
physical system (the REO phenomenon is not specific to any particular mass
range), but instead choose them for ease of comparison to \citet{blaes02},
and because we wish to study this phenomenon in the test-particle case.

At each choice of $e_2$ we calculate the merger time.  We define a merger
as occurring when the two components of the inner binary come within 10
Schwarzschild radii ($R_{\rm{Sch}}$) of each other (where $R_{\rm{Sch}}$
hereafter refers to the Schwarzschild radius of the larger BH).  We are
forced to integrate only to 10 $R_{\rm{Sch}}$ rather than down to 1 or 2
$R_{\rm{Sch}}$ because the PN terms begin diverging when the relative
velocity exceeds $\sim$0.2$c$ \citep[see Section 9.6 of][]{blanchet06}.  In
the systems we examine the relative velocities start to become close to
$\sim$0.2$c$ when the two components come within 10 $R_{\rm{Sch}}$ of each
other.  In practice, when the inner objects come within 10 $R_{\rm{Sch}}$
of each other, the orbital decay timescale is short compared to the overall
merger time.\footnote{To verify this we reran the system displayed in
Figure \ref{fig:example} and set the merger criterion to 100, 50, 20, 10,
and 5 Schwarzschild radii.  The overall merger times are all within
$\sim$0.01\% of each other.}

\begin{figure*}
\centering
\includegraphics[width=16cm]{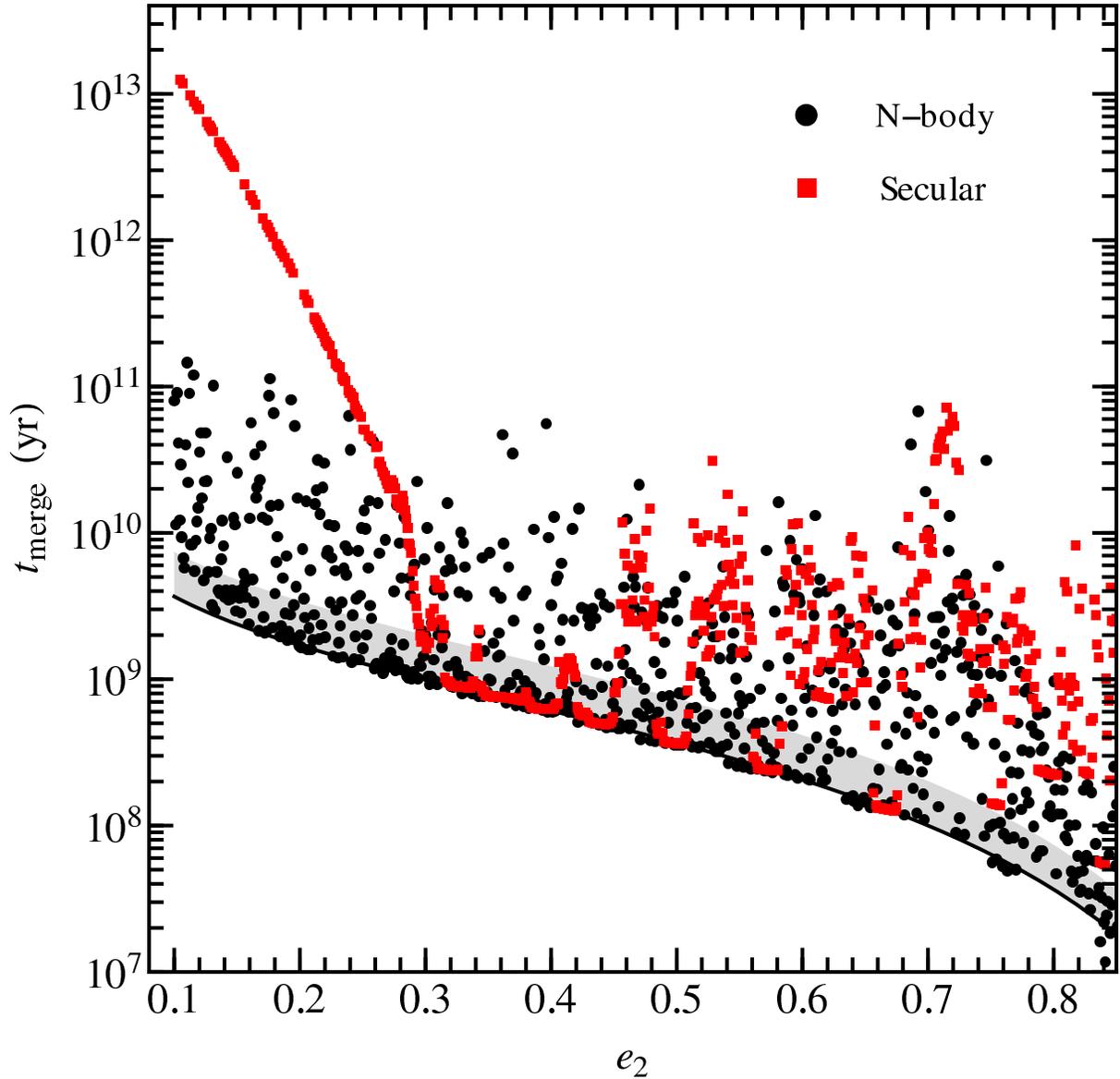}

\caption{The time required for the inner binary of triple systems to merge
as a function of the eccentricity of the orbit of the tertiary (see
Table~\ref{tbl:params} for the system parameters) using direct three-body
integrations (points) and using the secular approximation (red dashed
line).  The scatter in both the direct three-body integration and in the
double-orbit averaged calculation is due to the fact that these systems are
chaotic.  Slight changes in $e_2$ or the initial mean anomalies can change
$t_{\rm{merge}}$ by over an order of magnitude.  Approximately 25\% of the
systems we study merge in $t_{\rm{EKM}}$ (Eq.~\ref{eq:tekm}, solid line).
The shaded region depicts merger times within $1-2 \times t_{\rm{EKM}}$.
At $e_2 \lesssim 0.3$ the eccentric Kozai-Lidov mechanism weakens and
cannot drive systems to merger as shown by the large difference between the
secular and three-body calculations.  As a consequence, REOs become an
important mechanism to drive binaries to merger.  Because REOs are
fundamentally non-secular, the secular calculations overpredict the merger
times by many orders of magnitude at low $e_2$.} \label{fig:tmerge}

\end{figure*}

The results of these calculations appear in Figure~\ref{fig:tmerge}.
Because these orbits spend the vast majority of time in the Newtonian
regime, they can be rescaled to other masses, distances, and times until
the small fraction of time prior to merger that the eccentricity becomes
large enough that relativistic effects become important.  For this reason
we run each calculation to completion even if the merger time exceeds a
Hubble time for the particular case that we analyze.

There is substantial scatter in $t_{\rm{merge}}$.  This scatter is a result
of the slightly different choices of $e_2$ from point to point, but also
the different mean anomalies.  Two systems with identical starting
conditions but different initial mean anomalies can have merger times that
differ by up to two orders of magnitude.  The most rapidly merging systems
all merge within one eccentric Kozai-Lidov timescale.  This timescale is
given by \citet{katz11} and \citet{naoz13a} as
\[
t_{\textrm{EKM}} \sim \frac{t_{\textrm{KL}}}{\epsilon_{\textrm{oct}}},
\]
where $\epsilon_{\rm{oct}}$ is the strength of the octupole-order term in
the expansion of the three-body Hamiltonian.  The eccentric Kozai-Lidov
timescale can be written as
\begin{multline}
\label{eq:tekm}
t_{\textrm{EKM}} \sim 2.1 \times 10^9 \, \textrm{yr} \left( \frac{m_1 +
m_2}{2 \times 10^6 \, M_{\odot}} \right)^{-1/2} \left( \frac{a_1}{1 \,
\textrm{pc}} \right)^{3/2} \\
\times \left( \frac{m_1 + m_2}{2 m_3} \right) \left( \frac{a_2 / a_1}{20}
\right)^4 \frac{(1 - e_2^2)^{5/2}}{e_2}.
\end{multline}
This function matches the lower envelope of the merger time distribution
very closely.  Systems above this line fail to merge within a single
eccentric Kozai-Lidov cycle, but merge after several.  Usually, however,
the first eccentric Kozai-Lidov cycle so disturbs the system that future
eccentric Kozai-Lidov cycles operate on different timescales.  This is
primarily due to changes in the argument of periapsis of the inner binary.
The merger times are therefore not integer multiples of the first eccentric
Kozai-Lidov timescale.

Of the systems we study, approximately one-quarter merge within one
eccentric Kozai-Lidov cycle.  This is an overestimate of the true fraction
of systems that merge within $t_{\rm{EKM}}$ since we study only a small
region of parameter space.  Although our choice of inclination ($i =
80^{\circ}$) is not finely tuned, our choice of arguments of periapsis
($g_1 = 0^{\circ}, g_2 = 90^{\circ}$) is tuned to encourage strong
eccentric KL resonances.  We do this for two reasons.  First, it is
computationally expensive to integrate a sufficient number of systems to
marginalize over the arguments of periapsis and obtain good statistics.
Second, it demonstrates more clearly that the lower envelope of the
$t_{\rm{merge}}$ distribution is set by $t_{\rm{EKM}}$ because both
Kozai-Lidov resonances and REOs are stronger when the arguments of
periapsis are different by 90$^{\circ}$.

\begin{figure*}
\centering
\includegraphics[width=16cm]{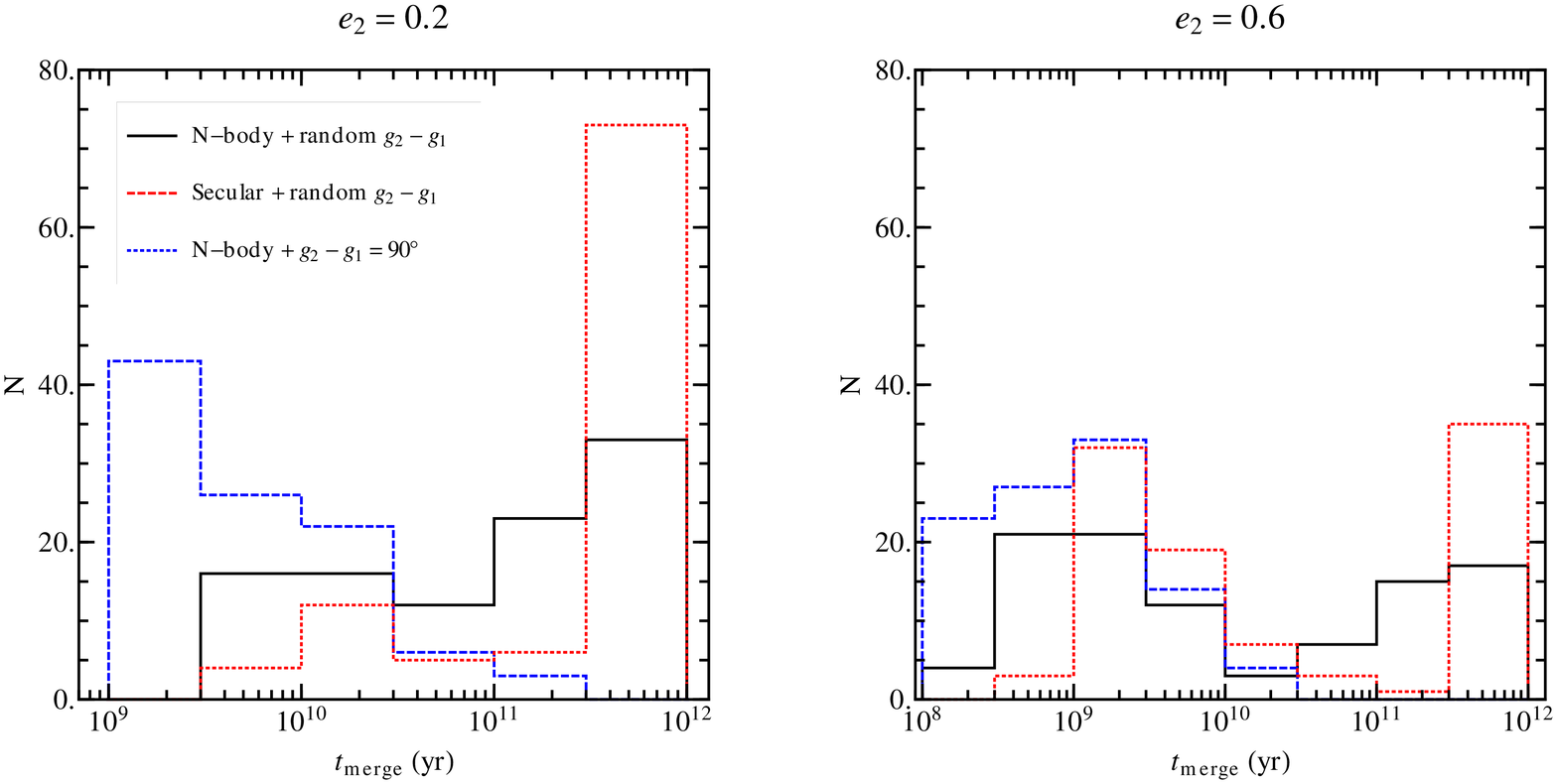}

\caption{Merger time distribution for fixed $e_2$.  We compare the
distribution when the arguments of periapsis are chosen randomly from a
uniform distribution (solid line) to when the arguments of periapsis are
fixed at $g_1 = 0^{\circ}$ and $g_2 = 90^{\circ}$ calculated using direct
three-body integration (blue dashed line) and in the secular approximation
with a uniform distribution of arguments of periapsis (red dotted line) for
100 systems.  Since Kozai-Lidov oscillations and REOs are stronger when the
arguments of periapsis differ by $90^{\circ}$, the merger time distribution
shifts to longer merger times when the arguments of periapsis are chosen
randomly.  About $15-30$\% of systems still merge rapidly when the
arguments of periapsis are chosen randomly.  The last bin of the secular
calculation is a lower bound.  The distribution is shifted to larger merger
times when the secular approximation is employed.} \label{fig:ghist}

\end{figure*}

To examine the effect of a more realistic distribution of initial arguments
of periapsis on the merger time distribution, we pick two choices of $e_2$
(0.2 and 0.6), and calculate the evolution of 100 systems with uniform
distributions of $g_1$ and $g_2$.  We compare this distribution of
$t_{\rm{merge}}$ (black line) to the distribution when the arguments of
periapsis are fixed to $g_1 = 0^{\circ}$ and $g_2 = 90^{\circ}$ (blue
dashed line) in Figure~\ref{fig:ghist}.  As expected, the distribution
shifts to longer merger times when the arguments of periapsis are chosen
randomly.  Nevertheless, about $\sim$15\% of systems with $e_2 = 0.2$ and
about $\sim$30\% of systems with $e_2 = 0.6$ merge within a few $\times \,
t_{\rm{EKM}}$.  We additionally compare this result to that calculated in
the secular approximation and find that it is shifted to much longer
$t_{\rm{merge}}$ than either calculation performed using direct three-body
integration.

Double-orbit averaging fails to correctly predict the merger times most
drastically for triple systems in which the tertiary is at low
eccentricity.  At best the secular calculation overpredicts the merger time
by two orders of magnitude, and at worst it overpredicts the merger time by
nearly four.  The catastrophic failure of double-orbit averaging is due to
the fact that it cannot account for REOs.  When the orbit of the tertiary
has a low eccentricity, Kozai-Lidov resonances (including eccentric
Kozai-Lidov resonances) are weakened.  Consequently, when the outer orbit
is at sufficiently low eccentricities, the Kozai-Lidov resonance is not
strong enough to drive the inner binary to merger on its own.  Kozai-Lidov
resonances nevertheless drive the inner binary to sufficiently high
eccentricities that REOs become important.  REOs drive the inner binary to
higher eccentricities, thereby causing relativistic effects to become much
more important.  In particular, gravitational wave radiation is much more
efficient when the inner binary is at higher eccentricities.

\section{Discussion and conclusions}
\label{sec:conclusions}

We have performed an exploration of a dynamical effect in hierarchical
triple systems that is not captured by secular double orbit averaging.  By
directly integrating the orbits of the three bodies and including
post-Newtonian terms up to order 3.5, we show that the eccentricity of the
inner binary oscillates on the timescale of the period of the outer binary
with amplitude given by Eqs.~\ref{eq:reo}~and~\ref{eq:reo_ecc} (see Figure
\ref{fig:example} and \citealt{ivanov05, bode13}).  During most of the
evolution of the triple system these oscillations are negligible and
secular calculations are valid.  However, when the eccentricity of the
inner binary is close to unity, fluctuations in the angular momentum of the
inner binary due to REOs become comparable to the total angular momentum in
the inner binary itself.  This is because the system spends more time at
higher $e_1$ (see Figure~\ref{fig:etimefrac}).  We find that the time spent
at eccentricities greater than any given eccentricity $e_1^{\prime}$ is
$\sim$few $\times \, (1 - e_1^{\prime})$.  As a consequence of this, REOs
can substantially affect the dynamics of the triple system.  Though we have
limited our analysis in this paper to triples of SMBHs for concreteness,
our results apply generally to any triple system for which the inner binary
consists of black holes or neutron stars.  Because relativistic effects are
extremely strong functions of eccentricity, REOs can drive binaries to
merge more rapidly by many orders of magnitude.  As this discrepancy occurs
over a broad range of parameter space, REOs will drive many systems to
merge which otherwise would not merge within a Hubble time.  Though a
complete treatment of the delay time distribution of compact object mergers
across a broad range of parameter space is beyond the scope of this paper,
REOs may be an important correction to calculations of the merger rate and
delay time distribution of compact object binaries in triple systems
\citep[e.g.,][]{thompson11, katz12} and perhaps systems like stars and
planets that may be strongly affected by tides (see Figure~\ref{fig:tmerge}
and Section \ref{subsec:tides} below).

\subsection{Implications for extreme-mass-ratio inspirals}

\begin{figure}
\centering
\includegraphics[width=8cm]{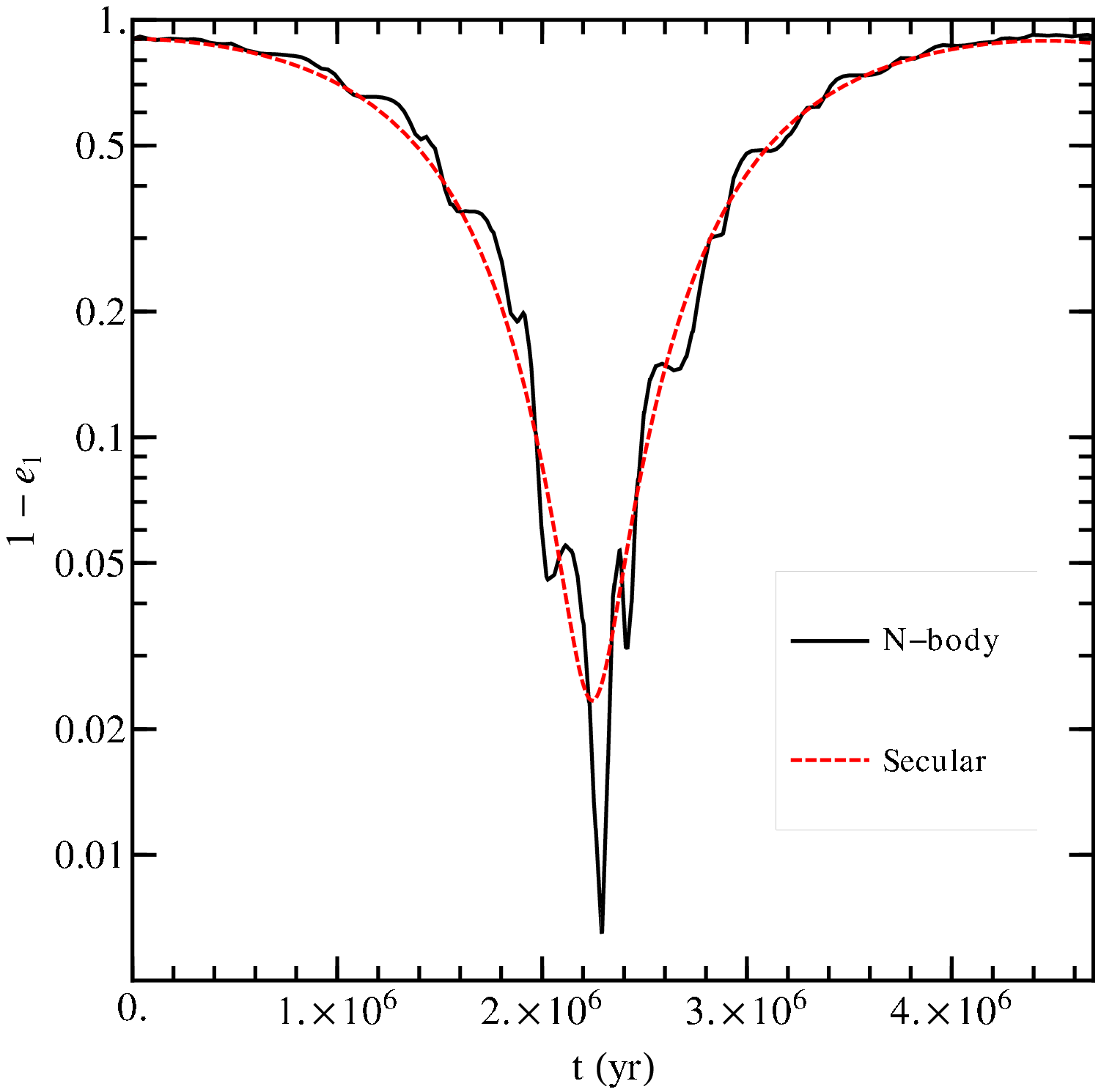}

\caption{REOs in an EMRI calculated with direct three-body integration
(solid line) and in the secular approximation (red dashed line).  Because
the mass ratio between the inner binary and outer binary is very large, the
amplitude of the fluctuations in the angular momentum in the inner binary
due to REOs becomes comparable to the angular momentum in the inner binary
itself.  Parameters of the system are provided in Table \ref{tbl:params},
but with $m_3 = 3 \times 10^8$ M$_{\odot}$, $g_1 = 0^{\circ}$, $g_2 =
90^{\circ}$, and $e_2 = 0.2$.}
\label{fig:emri}

\end{figure}

Since we have examined hierarchical triples with extreme mass ratios, a
potential application of our results is to extreme-mass-ratio inspirals
(EMRIs).  EMRIs consist of a binary of stellar-mass objects in orbit around
a SMBH \citep[see, e.g.,][]{amaro-seoane07, amaro-seoane12b,
amaro-seoane13b}.  In such cases there is an extremely large mass ratio
between the tertiary and the inner binary.  Although we ignore many
important features of real EMRIs (e.g., a stellar background, which leads
to several important phenomena, such as the Schwarzschild barrier,
discussed in \citealt{amaro-seoane12b}, but also the role of the spin of
the central MBH, \citealt{Amaro-SeoaneSopuertaFreitag2013}), we here
discuss the potential impacts of REOs on EMRIs as a motivation to future
works.  Because the amplitude of the eccentricity oscillations given by
Equation~\ref{eq:reo} is proportional to the mass ratio, arbitrarily large
mass ratios can lead to arbitrarily large eccentricity oscillations.  If
the oscillations are too large they can unbind the inner binary.  But
encounters at large enough distances that the binary system does not become
unbound could therefore lead to REOs with amplitudes comparable to the
amplitude of the Kozai-Lidov resonance itself.  Eccentricity oscillations
for a fiducial EMRI are shown in Figure~\ref{fig:emri}.  At the peak of the
Kozai-Lidov cycle the oscillations reduce the distance at periapsis of the
inner binary by over a factor of five.  Since the gravitational wave merger
timescale for a very eccentric orbit is proportional to $(1 - e^2)^{7/2}$
\citep{peters64}, this reduction in the distance at periapsis due to REOs
could lead to a significant reduction in the merger time and increase
gravitational wave luminosity if the dynamical features of realistic EMRIs
do not suppress this effect.  These results should be studied with more
detail in the context of secular effects in semi-Keplerian systems with
relativistic corrections, such as in the works of
\citet{merritt11, brem12}.

\subsection{Implications for gravitational wave emission}

In this subsection we discuss an important application of our findings that
will be soon expanded in a statistical study of the dynamics and the
implications for ground-based gravitational wave detectors such as Advanced
LIGO/VIRGO, along with the detailed description of the modification of the
integrator to incorporate relativistic effects \cite{amaro-seoane13c}.

It is uncertain what the dynamics immediately prior to merger will be in
systems dominated by REOs.  Although the orbit will have substantially
circularized by the time the two objects of the inner binary come within 10
$R_{\rm{Sch}}$ of each other, the orbit nevertheless retains a
non-negligible eccentricity ($e_1 \sim 0.1$) in most systems \citep{wen03,
gould11}.  Whether or not typical compact object binaries retain
non-negligible eccentricity immediately prior to merger is of key
importance to gravitational wave detectors like LIGO and LISA.  Because
these experiments require gravitational wave templates to find
gravitational wave signals in their data, accurate a priori predictions of
the waveform shapes are crucial for the success of these experiments.
Gravitational wave searches like LIGO have generally assumed that by the
time a merging binary is emitting gravitational waves at frequencies to
which they are sensitive, it has completely circularized.  But if a
substantial number of compact object binaries are driven to merger due to
Kozai-Lidov resonances, then the assumption of perfectly circular inspirals
will be mistaken.  Since gravitational radiation is a very strong function
of distance, even a modest residual eccentricity ($e_1 \gtrsim 0.1$) would
suffice to bury a gravitational wave signal in the data if a circular orbit
template is used \citep{brown10}.

\begin{figure*}
\centering
\includegraphics[width=16cm]{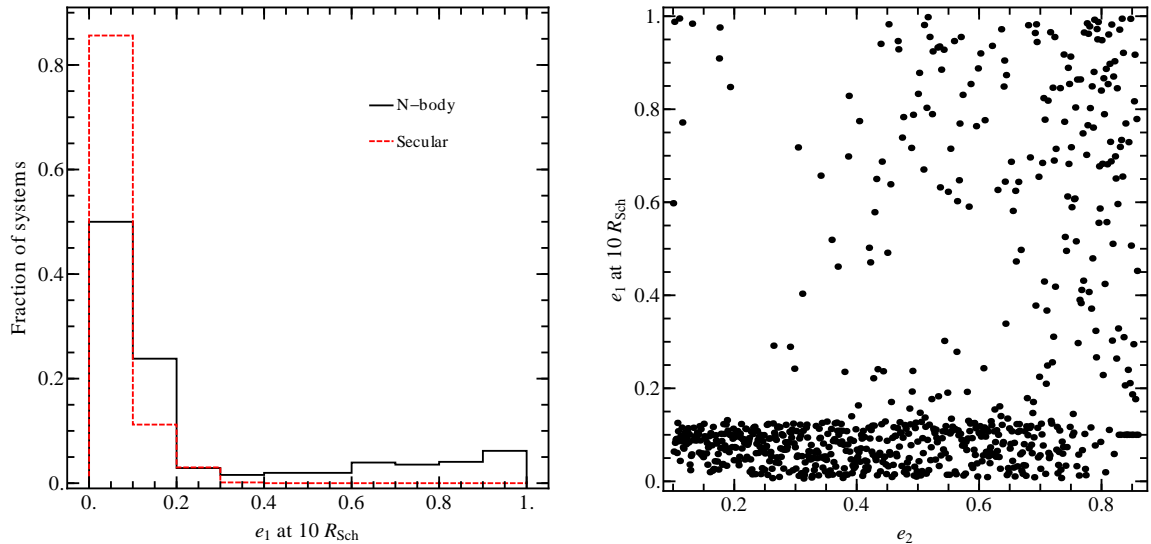}

\caption{The eccentricity distribution of the inner orbit when the inner
two components of the systems calculated in Figure~\ref{fig:tmerge} come
within 10 $R_{\rm{Sch}}$ of each other.  \emph{Left panel:} The secular
approximation (red dashed line) underpredicts the number of binaries which
merge at high eccentricities ($e \gtrsim 0.2$) relative to the direct
integration (black line).  In particular, the secular approximation
predicts that no binaries will come within 10 $R_{\rm{Sch}}$ at $e_1 >
0.4$, whereas the direct integration predicts that $\sim$20\% of
hierarchical triples do.  \emph{Right panel:} The distribution of
$e_{\rm{final}}$ as a function of $e_2$.  A larger fraction of systems
merge at high $e_{\rm{final}}$ when $e_2$ is large.  Note that we have
assumed a uniform distribution in $e_2$; if $e_2$ is distributed thermally
more systems will merge at high $e_1$.  Gravitational wave detectors will
need to employ templates of eccentric binaries to detect such systems.}
\label{fig:efin}

\end{figure*}

We calculate the eccentricity distribution of the inner binaries of the
triple systems investigated in Section \ref{subsec:population}.
Figure~\ref{fig:efin} presents the distribution of $e_1$ as the two
components of the inner binary come within 10 $R_{\rm{Sch}}$ of each other
calculated using direct three body integration (solid line) and in the
secular approximation (red dashed line).  At 10 $R_{\rm{Sch}}$ a Keplerian
orbit becomes an increasingly poor approximation to the true orbit of the
inner binary.  As such, we define the eccentricity of the orbit to be
\begin{equation}
\label{eq:eccentricity}
e_1 \equiv \sqrt{1 - \frac{1}{G (m_0 + m_1) a_1} \left( \frac{L_1}{\mu_1}
\right)^2},
\end{equation}
where $L_1$ is the angular momentum of the inner binary and $\mu_1$ is the
reduced mass of the inner binary.  We add PN corrections to $L_1$ up to
second order \citep[e.g.,][]{iyer95}.  To increase the computational
efficiency, in the secular approximation we calculate systems until $a_1$
has decreased to 1\% of its initial value.  At this point the inner binary
has decoupled from the outer binary and can be calculated independently.
We then calculate the eccentricity of the orbit when the two components
come within 10 $R_{\rm{Sch}}$ of each other using the adiabatic calculation
of \citet{peters64}.

The eccentricity distribution calculated using direct integration predicts
that $\sim$10\% of systems merge at high eccentricity ($e_1 > 0.8$).  In
the secular approximation, however, nearly all systems merge at low
eccentricity ($e_1 \lesssim 0.2$).  We also present the distribution of
$e_1$ at 10 $R_{\rm{Sch}}$ as a function of $e_2$ in the right panel of
Figure~\ref{fig:efin}.  Triples with larger $e_2$ have a greater chance of
merging at high eccentricity.  Since we assume a uniform distribution of
$e_2$ in the left panel of Figure~\ref{fig:efin}, a more realistic thermal
distribution will lead to a larger fraction of systems merging at high
eccentricity.  Population synthesis studies of hierarchical triple systems
which employ the secular approximation will therefore miss an important
source of unique gravitational waveforms.  If an important residual
eccentricity was indeed typical in these situations, gravitational wave
experiments would possibly have to take this into account in the
preparation of the waveform banks.  This question is addressed in detail in
\citet{amaro-seoane13c}.

One interesting possible outcome of a NS-NS merger in a triple system would
be a head-on collision similar to those between white dwarfs described in
\citet{katz12}.  In the case of binaries consisting of objects more compact
than white dwarfs, however, the chance of a collision is much smaller.
This is due to two factors.  Firstly, the objects themselves have a smaller
cross section than do white dwarfs by a factor of $\sim$10$^{6-7}$.
Secondly, close encounters will lead to circularization of the orbit at
larger distances relative to the object's radius due to the stronger
relativistic effects.  Head-on collisions between pairs of neutron stars or
black holes should therefore be rarer than collisions between WD-WD
binaries by a large factor.

\subsection{Tides and implications for stars and planets}
\label{subsec:tides}

Another well known class of triple systems with high mass ratios is that of
planets in binary star systems.  The formation of hot Jupiters is a
long-standing problem in the theory of planet formation and Kozai-Lidov
cycles have been proposed as a mechanism to drive planets formed far from
the host star into tight orbits \citep{wu07}.  Tidal effects are very
important for stellar and planetary systems and while a complete treatment
is beyond the scope of this paper \citep[though see][for a discussion of
the effect of tides on Kozai-Lidov cycles]{naoz12}, we nevertheless make
some qualitative statements about the impact of tides on our results and
the implications for stars and planets.

The overall effect of tides on eccentric orbits is to circularize them and
reduce the semi-major axis \citep{hut81} on a characteristic tidal friction
timescale $t_{\rm{TF}}$.  If the eccentricity is very close to unity,
$t_{\rm{TF}} \propto (1-e_1)^{-3/2}$ \citep{hut82}.  Tides therefore
prevent stars and planets from remaining on high eccentricity orbits for
long periods of time and will disrupt sufficiently strong Kozai-Lidov
cycles.  However, REOs occur on a shorter timescale than the Kozai-Lidov
cycle by a factor of $P_2/P_1$.  Tides may not have enough time to
circularize the orbit at a relatively low eccentricity before REOs drive
the orbit to higher eccentricities.  Because REOs can reduce $(1-e_1)$ by a
factor of $\sim$5, this results in a reduction in $t_{\rm{TF}}$ by an order
of magnitude.  The orbit will thus circularize more rapidly and be brought
into a closer orbit than it would by Kozai-Lidov oscillations calculated in
the secular approximation.

\section{Acknowledgements}

TAT thanks Nate Bode and Chris Wegg for bringing these rapid eccentricity
oscillations to his attention.  We would like to thank Smadar Naoz for
useful discussions and John Fregeau for releasing \textsc{Fewbody} under
the GNU General Public License.  PAS acknowledges the hospitality of the
Kavli Institute for Theoretical Physics where one part of this work has
been completed.  This research was supported in part by the National
Science Foundation under Grant No.~NSF PHY11-25915.  BJS was supported by a
Graduate Research Fellowship from the National Science Foundation.  This
work is supported in part by a grant from the National Science Foundation.
This work has been supported by the Transregio 7 ``Gravitational Wave
Astronomy'' financed by the Deutsche Forschungsgemeinschaft DFG (German
Research Foundation).

\bibliographystyle{mn2e}
\bibliography{refs}

\begin{appendix}

\section{Details of the numerical methods}
\label{sec:appendix}

We have extended \textsc{Fewbody} to include post-Newtonian (PN) force
terms up to order 3.5, presently the state of the art.  Due to their
length, we do not reproduce the terms here (the third-order term alone
spans more than a page), but instead refer the reader to Equations~182,
183, 185, and 186 of \citet{blanchet06}.  These terms are conjectured to
accurately reproduce general relativistic effects to within several
Schwarzschild radii.  Though analytic error estimates do not exist in the
literature, comparisons of PN calculations with direct integration of the
Einstein field equations support the consensus that the PN terms are
effective to within this range \citep{will11}.

The PN terms are stronger functions of velocity and radius than the
Newtonian term.  Consequently, the inclusion of the PN terms makes
integration of the orbits much more difficult, particularly for highly
eccentric orbits where the radial distance and velocity are both changing
very rapidly.  Thus, while the PN terms might in principle be accurate down
to several Schwarzschild radii, in practice the efficient computation of
the orbit may limit the regime of applicability.

These difficulties are compounded by the roundoff error introduced by
integrating close encounters far from the origin \citep[see, e.g.,][for
further discussion]{mikkola08}.  If the positions of two nearby objects are
represented with respect to a distant origin, the numerical precision is
reduced by roughly the ratio of the distance of the two objects from the
origin to their separation.  (For example, if a computer has only four
digits of precision and two objects are separated by $1.234 \times 10^{-3}$
and are a distance of 1 from the origin, their positions must be
represented by 1.001, leading to a loss of three digits.)  In practice this
can lead to a loss of six or seven orders of magnitude of precision and can
render the evolution of high eccentricity systems intractable.  In general,
roundoff error in $N$-body dynamical simulations can be avoided with some
variation of algorithmic regularization \citep[see, e.g.,][]{mikkola99,
aarseth03}.  However, because we are only concerned with the special case
of hierarchical triple systems, we have modified \textsc{Fewbody} to avoid
roundoff error by automatically recentering the triple on every step so
that the center of mass of the inner binary is placed at the origin.

To test the correctness of our implementation and to characterize its
regime of accuracy we run several numerical tests.  We first examine the
degree of energy conservation in orbits at several eccentricities when no
PN terms are included and when non-radiative PN terms are included.  We
then show that the orbital decay due to the 2.5 order PN term closely
matches analytic calculations.  Finally we compare the results from
\textsc{Fewbody} to an octupole-order secular model in several simple cases
to demonstrate that systems in which the approximations of the secular
model are valid produce similar behavior as direct three-body integration.

\subsection{Energy conservation}
\label{subsec:encons}

The most straightforward way to determine the numerical accuracy of an
$N$-body integrator is to determine how well energy is conserved.  Once the
change in energy becomes non-negligible compared to the total energy it is
certain that the calculated dynamics are qualitatively incorrect.  Typical
energy conservation tolerances are set at least several orders of magnitude
below this point.  The largest tolerance often invoked is of order
$10^{-5}$.

In ordinary integration (i.e., without taking a Kustannheimo-Stiefel or
similar transformation), energy conservation is worst during close
encounters of extremely eccentric orbits.  Due to the steep $1/r^2$ profile
of the gravitational force and the rapid change in $r$ near the periapsis
of a highly eccentric orbit, such orbits are difficult to calculate
accurately.  This problem is exacerbated with the introduction of PN terms
since the PN terms are even stronger functions of distance and include
strong velocity terms which vary rapidly as well.  For an orbit with a
given semi-major axis, there is thus a maximum eccentricity to which we can
accurately integrate.

To estimate \textsc{Fewbody}'s numerical accuracy and determine this
maximum eccentricity, we integrate $3 \times 10^6$ orbits (approximately
one Hubble time) of two $10^7$ M$_\odot$ SMBHs with a semi-major axis of 1
pc and eccentricities ranging from $1-e = 1$ to $10^{-5}$.  This system is
the inner binary of the systems we integrate in \S \ref{sec:mergetime}.
Since the triple systems we later integrate consist of a tertiary with a
mass of $10^7$ M$_\odot$ at a distance of 20 pc, we offset the binary in
these energy tests to a distance of 10 pc so as to account for roundoff
error that will be present when we integrate the triple systems.  This
initial offset has only a negligible effect on the calculation, however,
because our code automatically recenters the triple system on the center of
mass of the inner binary on every step so as to eliminate this roundoff
error.

We perform these calculations both with and without the PN terms.  In
calculations with the PN terms we exclude the odd-order 2.5 and 3.5 terms
since these serve to describe the effects of gravitational radiation.
These force terms are not conservative and are therefore not appropriate in
our checks for energy conservation.  The accuracy of these odd terms is
characterized in \S \ref{subsec:inspiral}.  We further note that when PN
force terms are included in the integration, the expression for the energy
changes.  The energy term including PN terms up to third-order is lengthy,
so like the force terms, we do not reproduce it here, but instead refer the
reader to Equation~2.11 of \citet{mora04}.  \citep[The energy term is also
provided in Equation~170 of][but is represented in a different gauge that
contains an undesirable logarithm.]{blanchet06}

\begin{figure}
\centering
\includegraphics[width=8cm]{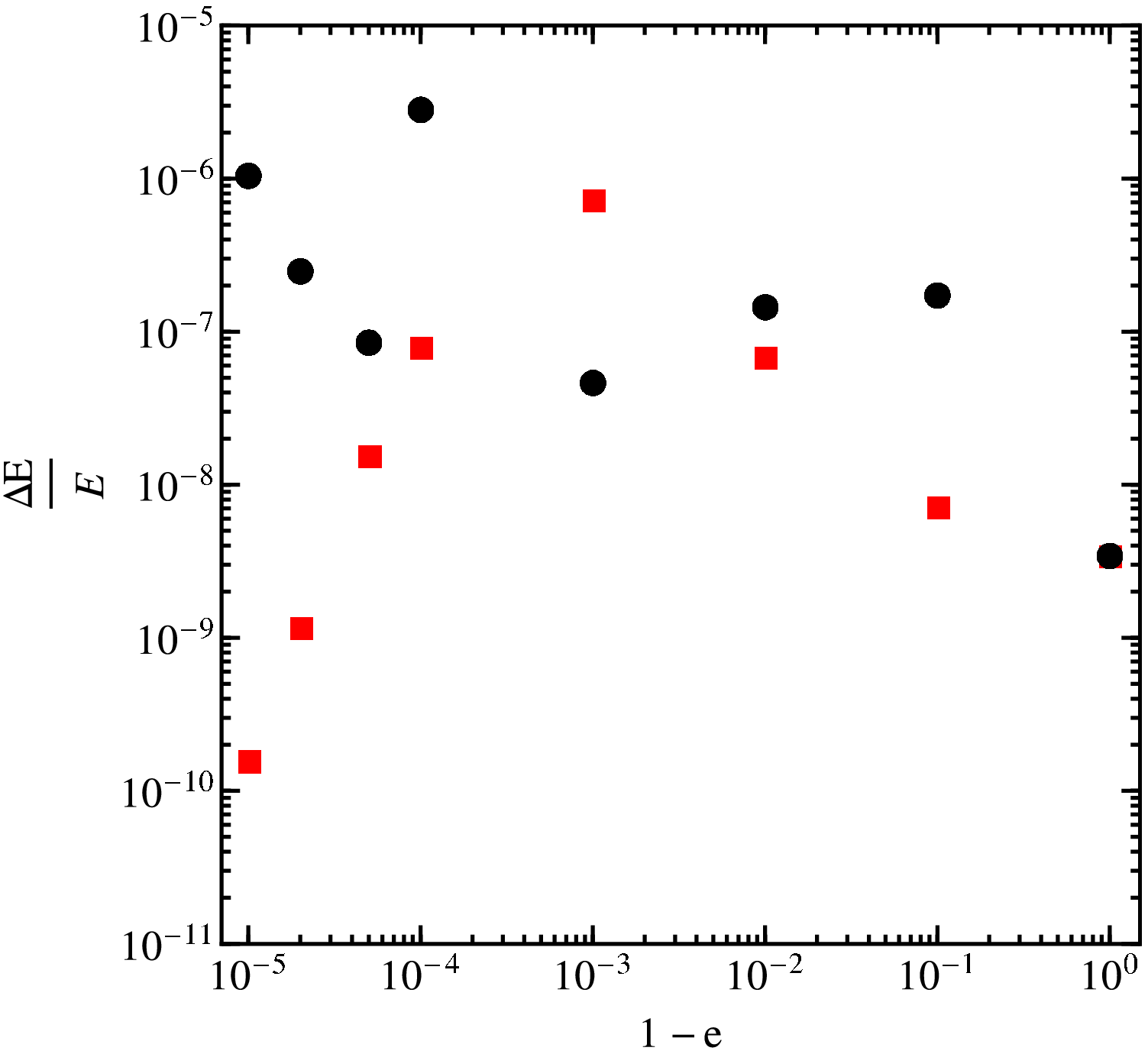}

\caption{Energy conservation in \textsc{Fewbody} for orbits over a range of
eccentricities in the Newtonian case (red squares) and including
non-radiative post-Newtonian terms up to order 3 (black dots).  We
integrate the orbits for a Hubble time or for the gravitational radiation
inspiral time, whichever is less.  In all cases energy conservation is
better than $10^{-5}$.}\label{fig:encons}

\end{figure}

For the low-eccentricity systems ($1-e \geq 10^{-3}$), gravitational
radiation is weak and so we integrate them for a Hubble time.  The
high-eccentricity systems will merge in under a Hubble time, however, so we
simply integrate them for as long as it takes them to merge via
gravitational radiation.  The results of these integrations are presented
in Figure~\ref{fig:encons}.  The results from including the PN 1 term alone
and from including just the PN 1 and PN 2 terms are very close to the
results from including all three PN terms.  They are therefore not
displayed in Figure~\ref{fig:encons}.  As expected, \textsc{Fewbody}
conserves energy best at low eccentricities.  At high eccentricities energy
conservation is also quite good because only a small number of orbits need
to be integrated.  At all eccentricities, however, \textsc{Fewbody}
performs the integration for the necessary number of orbits and conserves
energy to well under $10^{-5}$ (and often much better).

\subsection{Inspiral time}
\label{subsec:inspiral}

To test the accuracy of the radiation reaction terms we compare the orbital
decay of a highly eccentric orbit to the analytic expressions of
\citet{peters64}.   From \citet{peters64}, the semi-major axis and
eccentricity evolution of the orbit are described by the following two
differential equations:
\[
\left< \frac{da}{dt} \right> = - \frac{64}{5} \frac{G^3 m_1 m_2 (m_1 +
m_2)}{c^5 a^3 (1 - e^2)^{7/2}} \left( 1 + \frac{73}{24} e^2 + \frac{37}{96}
e^4 \right),
\]
\[
\left< \frac{de}{dt} \right> = - \frac{304}{15} e \frac{G^3 m_1 m_2 (m_1 +
m_2)}{c^5 a^4 (1 - e^2)^{5/2}} \left(1 + \frac{121}{304} e^2 \right).
\]
The orbital decay time of the system is given by the integral
\[
T(a_0, e_0) = \frac{12}{19} \frac{c_0^4}{\beta} \int_0^{e_0}
\frac{e^{29/19} \left[1 + (121/304) e^2 \right]^{1181/2299}}{(1 -
e^2)^{3/2}} \rm{d}e,
\]
where
\[
\beta = \frac{64}{5} \frac{G^3 m_1 m_2 (m_1 + m_2)}{c^5},
\]
and
\[
c_0 = \frac{a_0 (1 - e_0^2)}{e^{12/19}} \left(1 + \frac{121}{304} e^2
\right)^{-\frac{870}{2299}}.
\]

To compare \textsc{Fewbody} to the analytic results, we calculate the
evolution of the orbital parameters of the orbit of two $10^7$ M$_{\odot}$
SMBHs with a semi-major axis of one parsec and an initial eccentricity of
$1-e = 10^{-4}$.  For the purposes of this comparison, we perform the
\textsc{Fewbody} calculation with the 2.5 PN term alone.  This is because
\citet{peters64} assumes that the orbits are Keplerian and calculates the
gravitational wave power in the quadrupole approximation.  For very
eccentric orbits, the deviation from a perfect ellipse manifests itself as
a longer dwelling time at periapsis.  Since most of the gravitational
radiation is emitted near periapsis, a fully relativistic orbit results in
more radiation emitted than \citet{peters64} predicts.  As the 2.5 PN term
is the only term that captures quadrupole radiation emission, this is the
only term consistent with the assumptions of \citet{peters64}.

We find that the difference in the overall merger time between
\textsc{Fewbody} and \citet{peters64} is less than $10^{-3}$.  We believe
this discrepancy is due to the fact that \textsc{Fewbody} treats the energy
loss more realistically by emitting most of the orbital energy during
passage through periapsis.  \citet{peters64}, however, assumes that energy
loss is continuous throughout the orbit.  For very eccentric orbits like
the ones we are modelling, \textsc{Fewbody}'s treatment leads to stepwise
changes in the orbital parameters, whereas \citet{peters64} assumes that
these orbital parameters vary continuously across the entire orbit.  Over
many orbits, this difference manifests itself in small discrepancies in the
orbital parameters between the two calculations.

During most of the orbit the semi-major axis calculated by \textsc{Fewbody}
is within $10^{-2}$ of the semi-major axis predicted by \citet{peters64}.
At the end of the inspiral the discrepancy is much worse, but this is
simply because the overall merger time of the orbit is slightly different
between \textsc{Fewbody} and \citet{peters64}.  Although we therefore
cannot adequately test \textsc{Fewbody} in this regime, however, we are not
interested in the precise dynamics prior to merger, only the overall merger
time.

\begin{figure}
\includegraphics[width=8cm]{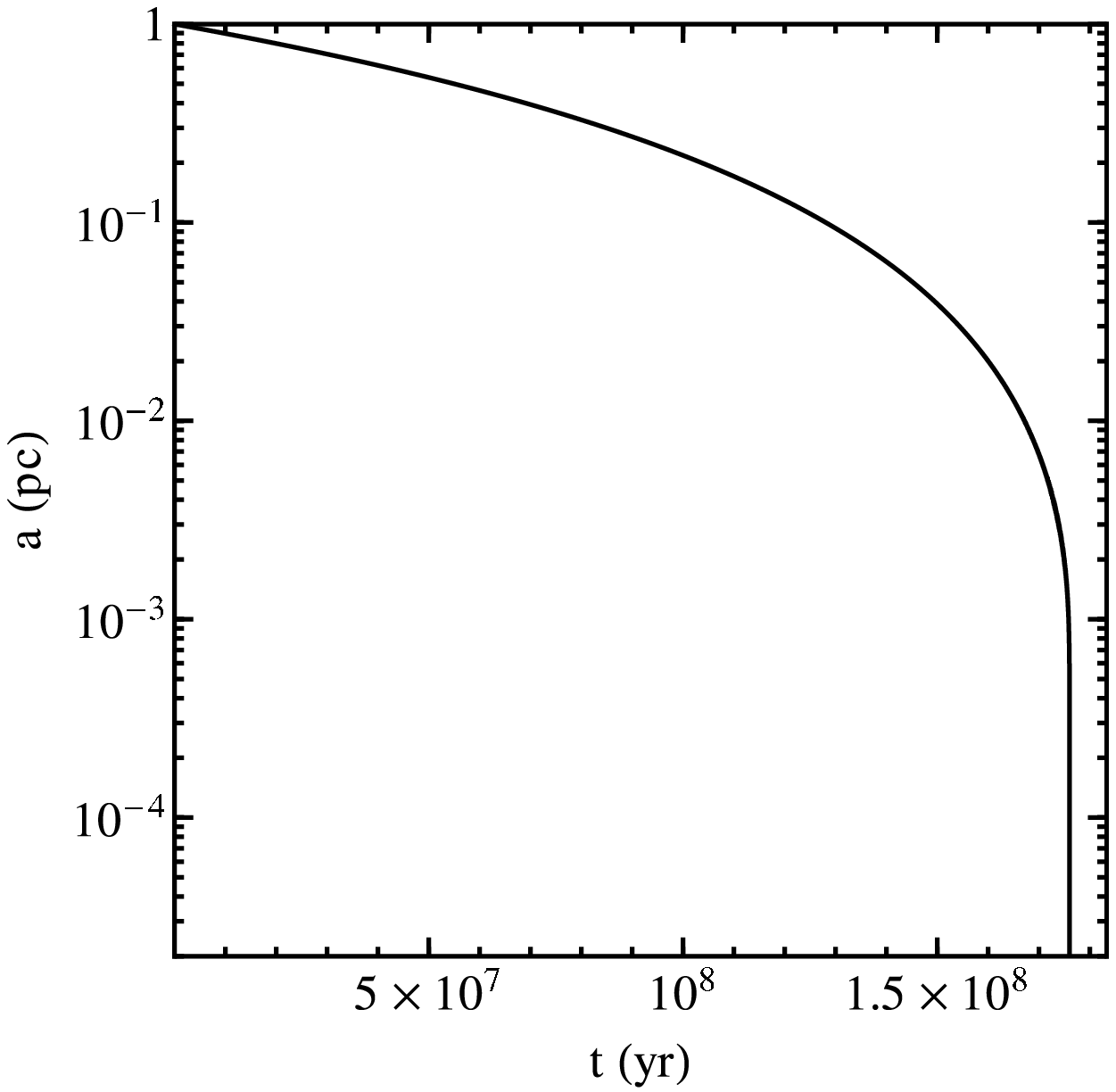}

\caption{The evolution of the semi-major axis of a binary of two $10^7$
M$_{\odot}$ SMBHs with an initial semi-major axis of 1 pc and an initial
eccentricity of $1-e \sim 10^{-3}$.  For purposes of comparison with
\citet{peters64}, the calculation shown includes only the 2.5 PN term.  The
difference between our results and \citet{peters64} is much less than the
thickness of the line.  We also calculate the evolution using all PN terms
up to and including PN 3.5.  In this experiment the difference in the
merger time between this calculation and the full PN calculation is 0.5\%.
For clarity we omit displaying the evolution in the full PN approximation.
At higher initial eccentricities the discrepancy between the full PN
calculation and the 2.5 PN term alone is larger.  The inner binaries that
we examine in this paper begin to suffer copious energy loss due to
gravitational radiation at an eccentricity of $1-e \sim 10^{-4}$.  At this
eccentricity the difference between the full PN calculation and the 2.5 PN
approximation in \citet{peters64} is $\sim$5\%.} \label{fig:inspiral}

\end{figure}

The orbital decay including only the 2.5 PN term is presented in
Figure~\ref{fig:inspiral}.  The effect of adding the additional PN terms is
a 0.5\% change in the overall merger time.  At higher eccentricities the
other PN terms become stronger and yield even larger discrepancies.

\subsection{Comparison to the secular approximation}
\label{subsec:secular}

If any changes to the orbital parameters in a hierarchical triple are slow
compared to the outer orbital period, the orbits need not be integrated
directly, but instead can be calculated from a time-averaged Hamiltonian
\citep[e.g.,][]{blaes02}.  Although we show in this paper that this
approximation breaks down in important regions of parameter space, there
are broad regimes in which this approximation works well.  In particular,
the secular approximation works very well when the Kozai-Lidov mechanism
does not excite extremely high eccentricities.

We here show the agreement between the orbital evolution in the secular
approximation with the direct three-body integration.  For the secular
approximation we use the formalism of \citet{blaes02}, which is an
octupole-order calculation that includes general relativistic precession
and gravitational radiation.  (Note that \citealt{blaes02} use the
equations of general relativistic orbital decay from \citealt{peters64}.
As discussed in \S \ref{subsec:inspiral}, this slightly underpredicts the
merger time for highly eccentric orbits.)  The orbital evolution is
compared to the explicit orbit integration using \textsc{Fewbody} for a
slowly-varying hierarchical triple undergoing Kozai-Lidov oscillations.
Because the Hamiltonian in \citet{blaes02} uses the results from
\citet{peters64} to account for gravitational radiation, the only radiation
term we include is PN 2.5.  Similarly, the formalism for handling apsidal
precession in \citet{blaes02} is equivalent to the first PN term.  Thus the
only PN terms we include in this comparison are PN 1 and PN 2.5.

\begin{figure}
\centering
\includegraphics[width=8cm]{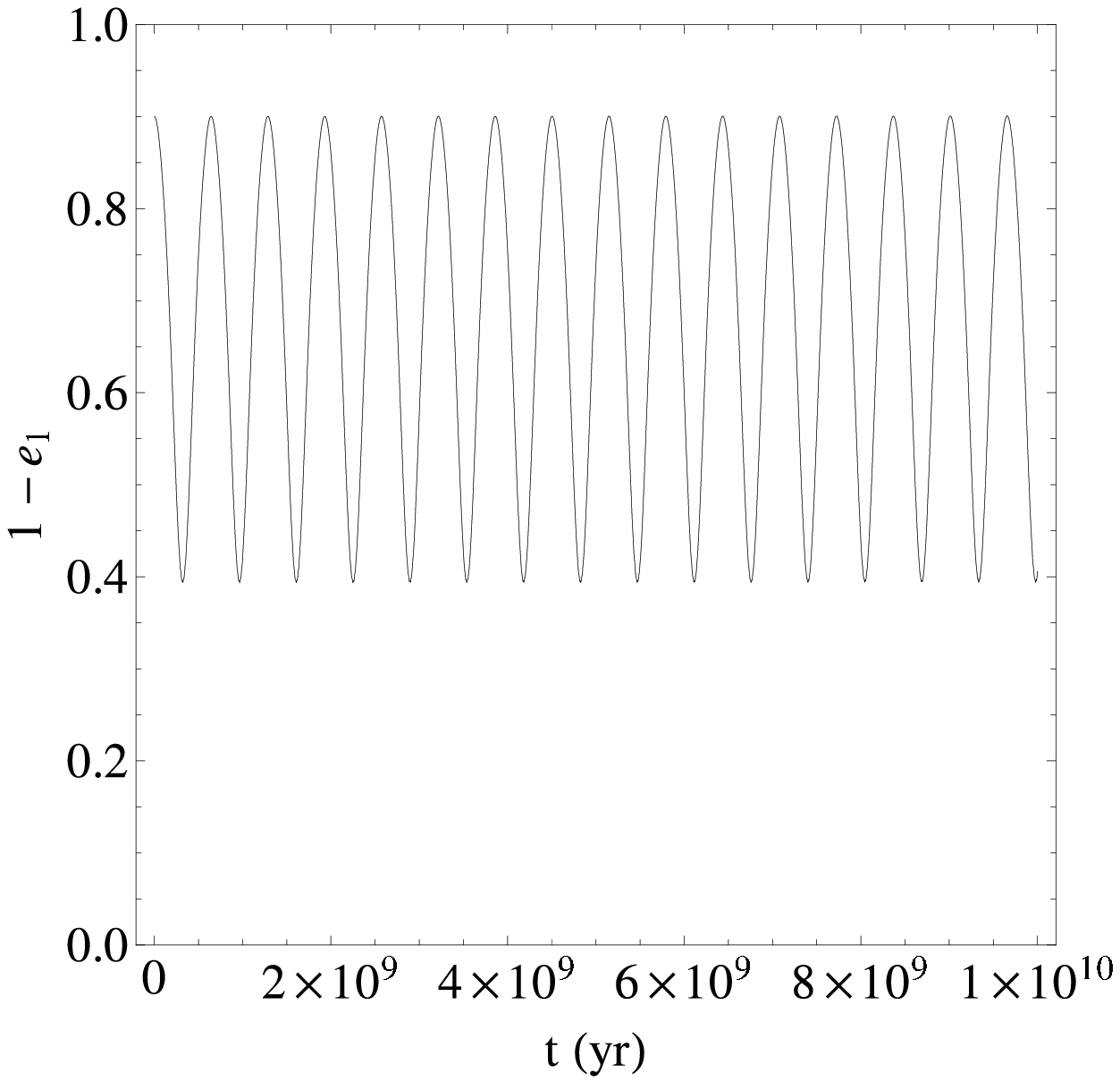}

\caption{The evolution in eccentricity of a system undergoing Kozai-Lidov
oscillations.  This system does not exhibit oscillations to extremely high
eccentricities, so it is in the regime in which the secular approximation
is valid.  Properties of this system are listed in Table
\ref{tbl:comparison}.  We evolved this system for $10^{10}$ years using
both the secular model of \citet{blaes02} and by performing the direct
three-body integration using \textsc{Fewbody}.  The difference between the
two techniques is much less than the thickness of the line and amounts to a
$\sim$0.1\% offset in time at the end of the calculation.  The difference
between the calculation in the secular approximation and the direct
three-body integration is explored further in
Figure~\ref{fig:phase}.}\label{fig:comparison}

\end{figure}

We calculate the evolution of a triple system in which the Kozai-Lidov
mechanism is present, but does not excite extremely high eccentricities.
Properties of this system are listed in Table \ref{tbl:comparison}.  We
evolve the system for $10^{10}$ yr, or about 15.5 Kozai-Lidov cycles.  The
eccentricity evolution of both calculations are presented in
Figure~\ref{fig:comparison}.  The difference between the two systems is
equivalent to a 0.1\% scaling in time.  This small difference is due to the
fact that \textsc{Fewbody} begins the integration with each object at a
random point along its orbit and at random longitudes of ascending node.
These different starting conditions yield slightly different orbits.  The
initial phases of the orbits and longitudes of ascending node do not impact
the orbit-averaged evolution of \citet{blaes02}.  The variation due to
these random initial conditions from one realization to the next is
consistent with the difference between any particular realization and the
orbit-averaged calculation.

\begin{table}
\centering

\caption{Initial conditions for a system that undergoes weak Kozai-Lidov
oscillations.  See Figure~\ref{fig:comparison} for the evolution of this
system.}\label{tbl:comparison}

\begin{tabular}{cc}
\hline
Parameter & Value \\
\hline

$m_1$ & $10^7$ M$_{\odot}$ \\
$m_2$ & $10^7$ M$_{\odot}$ \\
$m_3$ & $3 \times 10^6$ M$_{\odot}$ \\
$a_1$ & 1 pc \\
$a_2$ & 20 pc \\
$e_1$ & 0.1 \\
$e_2$ & 0.2 \\
$g_1$ & 0 \\
$g_2$ & 0 \\
$\cos i$ & 0.5 \\

\hline

\end{tabular}
\end{table}

\begin{figure}
\centering
\includegraphics[width=8cm]{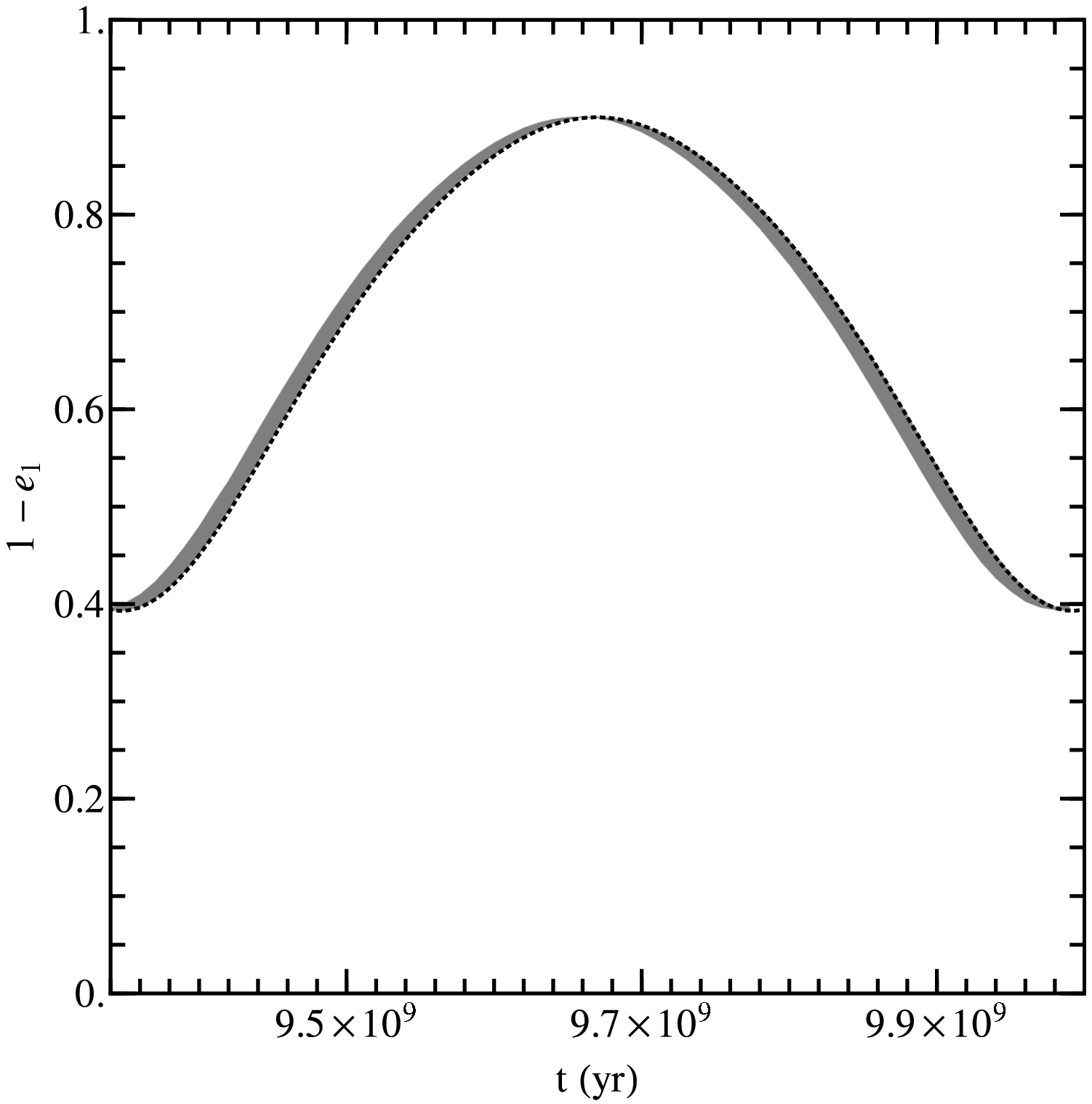}

\caption{The final moments of the evolution in eccentricity of the system
presented in Figure~\ref{fig:comparison}.  Properties of this system are
listed in Table \ref{tbl:comparison}.  The secular calculation (dotted
line) is shown with the results of 100 runs using \textsc{Fewbody} (gray
region).  The orbits of the triple system in each run were given random
initial mean anomalies.  The variation in the orbital evolution due to
these random initial mean anomalies results in a $\sim$0.15\% offset in
time.  The difference between the secular calculation and any given
calculation using \textsc{Fewbody} is consistent with this variation.}
\label{fig:phase}

\end{figure}

To more clearly illustrate the differences between the secular and
\textsc{Fewbody} calculations, we run the \textsc{Fewbody} calculation 100
times with random initial mean anomalies for each run.  The variation in
the evolution of the eccentricity of the inner orbit is shown in
Figure~\ref{fig:phase}.  The magnitude of this variation is a $\sim$0.15\%
scaling in the time, which amounts to an offset of $\sim$15 Myr after
$10^{10}$ yr.  The evolution predicted by the secular calculation is
consistent with the range calculated by \textsc{Fewbody}.

\end{appendix}

\end{document}